\documentclass[aps,prb,twocolumn,showpacs,superscriptaddress]{revtex4}

\usepackage{graphicx} 

\begin{document}

\title
{Metal nanoring and tube formation on carbon nanotubes}

\author{V. M. K. Bagci}
\affiliation{Department of Physics, Bilkent University, Ankara
06533, Turkey}
\author{O. G\"{u}lseren}
\affiliation{NIST Center for Neutron Research, National Institute
of Standards and Technology, Gaithersburg, Maryland 20899}
\affiliation{Department of Materials Science and Engineering,
University of Pennsylvania, Philadelphia, Pennsylvania 19104}
\author{T. Yildirim}
\affiliation{NIST Center for Neutron Research, National Institute
of Standards and Technology, Gaithersburg, Maryland 20899}
\author{Z. Gedik}
\affiliation{Department of Physics, Bilkent University, Ankara
06533, Turkey}
\author{S. Ciraci}
\affiliation{Department of Physics, Bilkent University, Ankara
06533, Turkey}

\date{\today}

\begin{abstract}

The structural and electronic properties of aluminum covered single
wall carbon nanotubes (SWNT) are studied from first-principles for
a large number of coverage. Aluminum-aluminum interaction that is
stronger than aluminum-tube interaction, prevents uniform metal
coverage, and hence gives rise to the clustering. However, a
stable aluminum ring and aluminum nanotube with well defined
patterns can also form around the semiconducting SWNT and
lead to metallization. The persistent current in the Al nanoring
is discussed to show that a high magnetic field can be
induced at the center of SWNT.

\end{abstract}



\maketitle

\section{Introduction}

Stable metal wires having diameters in the range of nanometer are
very important for nanoelectronics and other nanodevice
applications. Metal nanowires~\cite{agrait1,condrev} and
monoatomic chains~\cite{ohnishi,yanson} produced so far have
played a crucial role in understanding quantum transport and
exotic atomic structure.\cite{mehrez,oguz1,agrait2,tosatti}
Earlier, those wires were not reproducible and controllable to
offer any relevant technological application. Recently, it has
been shown that such nanowires can be produced by depositing metal
atoms on carbon nanotube templates.\cite{zhang1,zhang2} Because of
its curvature the surface of a single wall carbon nanotube (SWNT)
is chemically more reactive than graphite. Therefore stable bonding
can occur between the SWNT and metal
adatom.\cite{oguz2,deepak,oguz3} Recently, Mo-Ge superconducting
nanowires were fabricated using sputter deposition on carbon
nanotubes \cite{thinkham}. Continuous titanium coating of varying
thickness, and quasi continuous coating of Ni and Pd were obtained
by using electron-beam evaporation techniques,\cite{zhang1,zhang2}
whereas Au, Al, Fe, Pd were able to form only discrete particles or
clusters rather than a continuous coating of the SWNT.
Nevertheless, coating of virtually any metal on SWNT can be
mediated by depositing first titanium as a buffer
layer.\cite{zhang1,zhang2,ckyang}

SWNTs seem to be ideal templates for synthesizing a variety of
stable nanowires with different diameter, thickness and length of
elemental metals. It is therefore important to have a good
understanding of metal-SWNT interactions and mechanism of metal
coverage. In this paper, we address this issue from first-principles
by studying structural and electronic properties of Al adsorption
starting from a single atom adsorption to monolayer coverage.
We find that the Al-Al interaction is relatively stronger than the
Al-SWNT interaction, yielding Al-cluster formation rather than a
uniform coating over the SWNT for most of the cases. However, we
discovered that stable Al nanoring and also Al tube can form at
well-defined and ordered positions over the $(8,0)$ SWNT.
Furthermore, we estimate that the current through the Al nanorings
can produce large magnetic fields at the tip of a nanotube. We hope
that these findings will shed light into the usage of nanotubes as
a template to grow metal nanowires with many novel properties.

\section{Method}

The first principles total energy and electronic structure
calculations have been performed using the pseudopotential plane
wave method~\cite{castep} within the generalized gradient
approximation (GGA)~\cite{gga}. A tetragonal supercell have been
used with lattice constants, $a_{sc} = b_{sc} \sim 22$ \AA{} and
$c_{sc}$, which is taken to be equal to the 1D lattice parameter,
$c$, of the tube. To minimize the adsorbate-adsorbate interaction,
some calculations are performed in longer supercells by taking
$c_{sc}=2 c$. We used ultra-soft pseudopotentials~\cite{usps}
for carbon and aluminum atoms and plane waves up to an energy
cutoff of 310 eV. Brillouin zone integrations are performed with
12-6 special {\bf k}-points. All atomic positions of adsorbate
and nanotube as well as $c$ are fully optimized.

\section{Results and discussion}

We first explored the possible adsorption sites for an individual
Al atom on a (8,0) nanotube; namely \textbf{H-sites} which are above
the center of hexagons, \textbf{Z-} and \textbf{B-sites} which are
above the zigzag and axial C-C bonds, respectively, and finally
\textbf{T-sites} which are on top of the carbon atoms. The binding
energy is obtained from the expression,
\begin{equation}
E_{b} = E_{T}[SWNT] + E_{T}[Al] - E_{T}[Al+SWNT]
\label{eq:bind}
\end{equation}
in terms of the total energies of the fully optimized bare nanotube
($E_{T}[SWNT]$), the atomic Al ($E_{T}[Al]$), and the Al adsorbed
nanotube ($E_{T}[Al+SWNT]$). All total energy calculations are
carried out in the same supercell with $c_{sc}=2 c$. According to
the above definition stable structures have positive binding
energies. We find that the bindings at the T-sites are unstable;
the adatoms move to the H-sites upon relaxation. The bindings at
the H-, Z- and B-sites are found to be stable with the C-Al distance
2.28, 2.30, and 2.25 \AA, respectively.  The corresponding binding
energies are 1.70, 1.54 and 1.60 eV, respectively.\cite{contro}
According to Mulliken analysis, $\sim$~0.7e is transferred to the
nanotube upon absorption of a single Al atom
and partially occupied electronic states occur in the
band gap. While the binding energy of Al is negligible on the
graphite surface, the curvature of the $(8,0)$ tube provides
significant binding interaction.\cite{oguz2,oguz3,taner}

\begin{figure}
\includegraphics[scale=0.35]{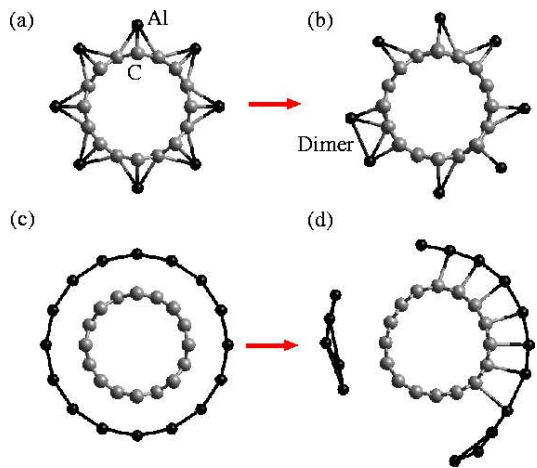}
\caption{ (a) Initial structure of the Al ring where the adatoms
were placed at the H-sites on the circumference of the tube,
(b) Dimerization upon relaxation of Al atoms starting from
structure shown in (a).
(c) Initial structure of uniform coverage of nanotube where
all H-sites are occupied by Al atoms. (d) The nucleation of
isolated Al clusters from the initial structure shown in (c).
Both relaxed structures shown in (b) and (d) are not final
equilibrium structures, but they are intermediate configurations
towards to 3D cluster formation.}
\label{fig:dimer}
\end{figure}

After having discussed the adsorption of a single Al atom
on a (8,0) nanotube, we next consider the adsorption of several
Al atoms where the Al-Al interactions play an important role.
Since the binding energy  at the H-sites is the largest,
we first consider a coverage where Al atoms are placed at the
H-sites. We start with a quarter coverage case (i.e. $\Theta=0.25$)
by placing eight Al atom at the H-sites around the circumference
forming a ring in the double unit cell of the $(8,0)$ nanotube.
According to this initial structure the Al-Al and C-Al distances
are 3.7, and 2.4 \AA, respectively. Once this system (consisting of
64 C and 8 Al atoms) is relaxed, we find that some Al atoms move
away from the nanotube surface towards their neighbors and
eventually form a dimer as shown in
Fig.~\ref{fig:dimer}(a$\rightarrow$b). We attribute this
dimerization to the  Al-Al interaction which is stronger
than the Al-nanotube interactions.

A uniform half coverage (i.e. $\Theta = 0.5$), where initially all
H-sites are occupied by Al atoms (\textit{i.e.} 32 C and 16 Al atoms)
exhibits also instability. Upon relaxation of this system, Al atoms
tend to reduce the Al-Al distance from 3.0 \AA{} to 2.5-2.8 \AA, and
at the same time they rise above the surface of the tube. At the end
small and isolated clusters form on nanotube. Some of the adatoms
become completely disconnected from the surface to initiate a three
dimensional (3D) island growth, since the latter is energetically
favorable with a binding energy $E_{b} > 3$ eV. This situation
as illustrated in Fig.~\ref{fig:dimer}(c$\rightarrow$d) is in
good agreement with the experimental
observations.\cite{zhang1,zhang2}

Above we showed that uniform Al coverage on the H-sites did not
yield a stable structure due to strong Al-Al interactions.
However, if the Al atoms are placed at the T-sites (i.e. on the
top of carbon atoms), the Al-Al distance no longer has to be large.
Therefore one can optimize the Al-Al and Al-nanotube interactions
simultaneously. Below we demonstrate this situation for two special
coverage cases, namely a zigzag Al-nanoring and a Al-nanotube around
the $(8,0)$ SWNT.

\begin{figure}
\includegraphics[scale=0.3]{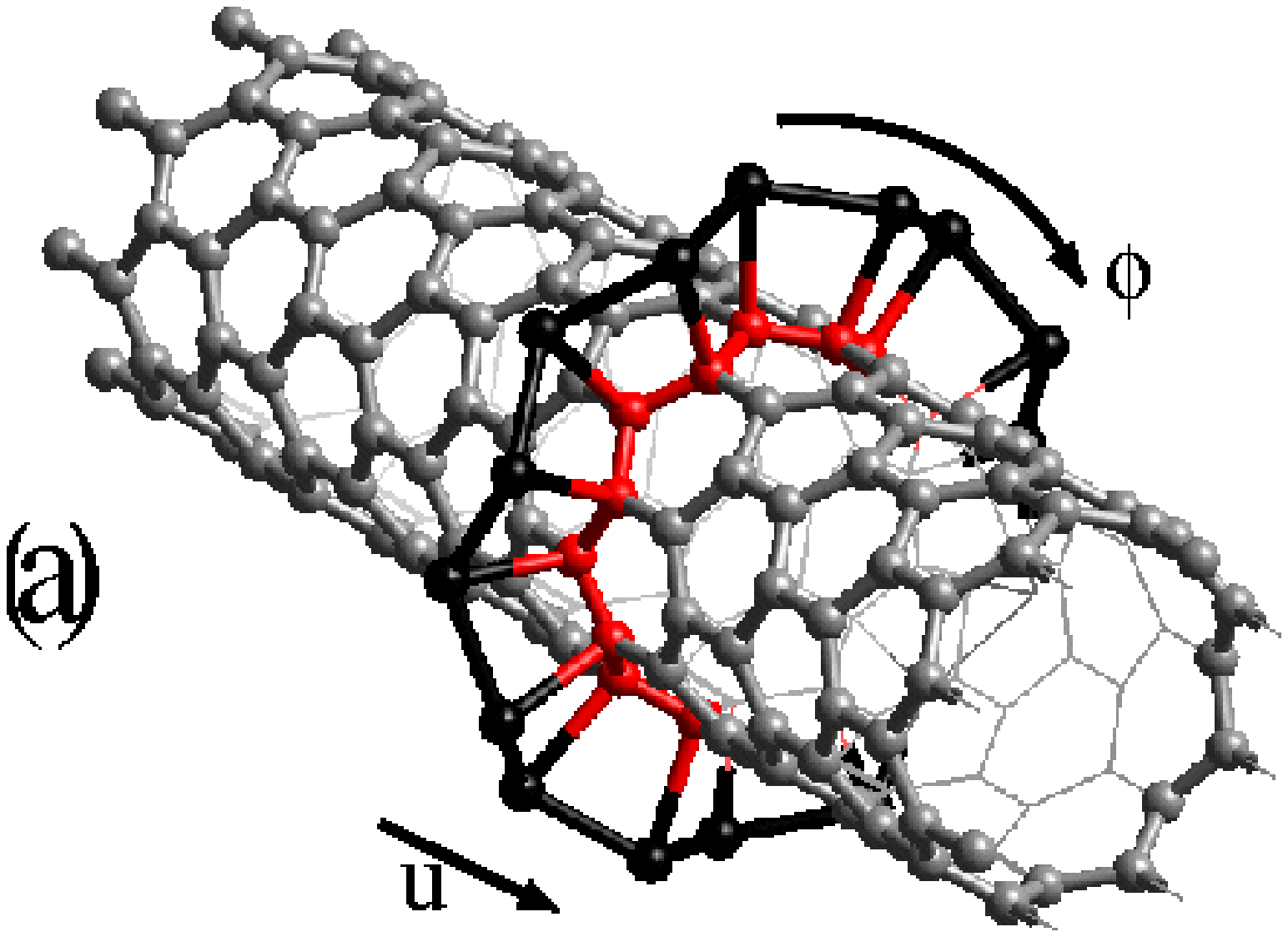}

\includegraphics[scale=0.2]{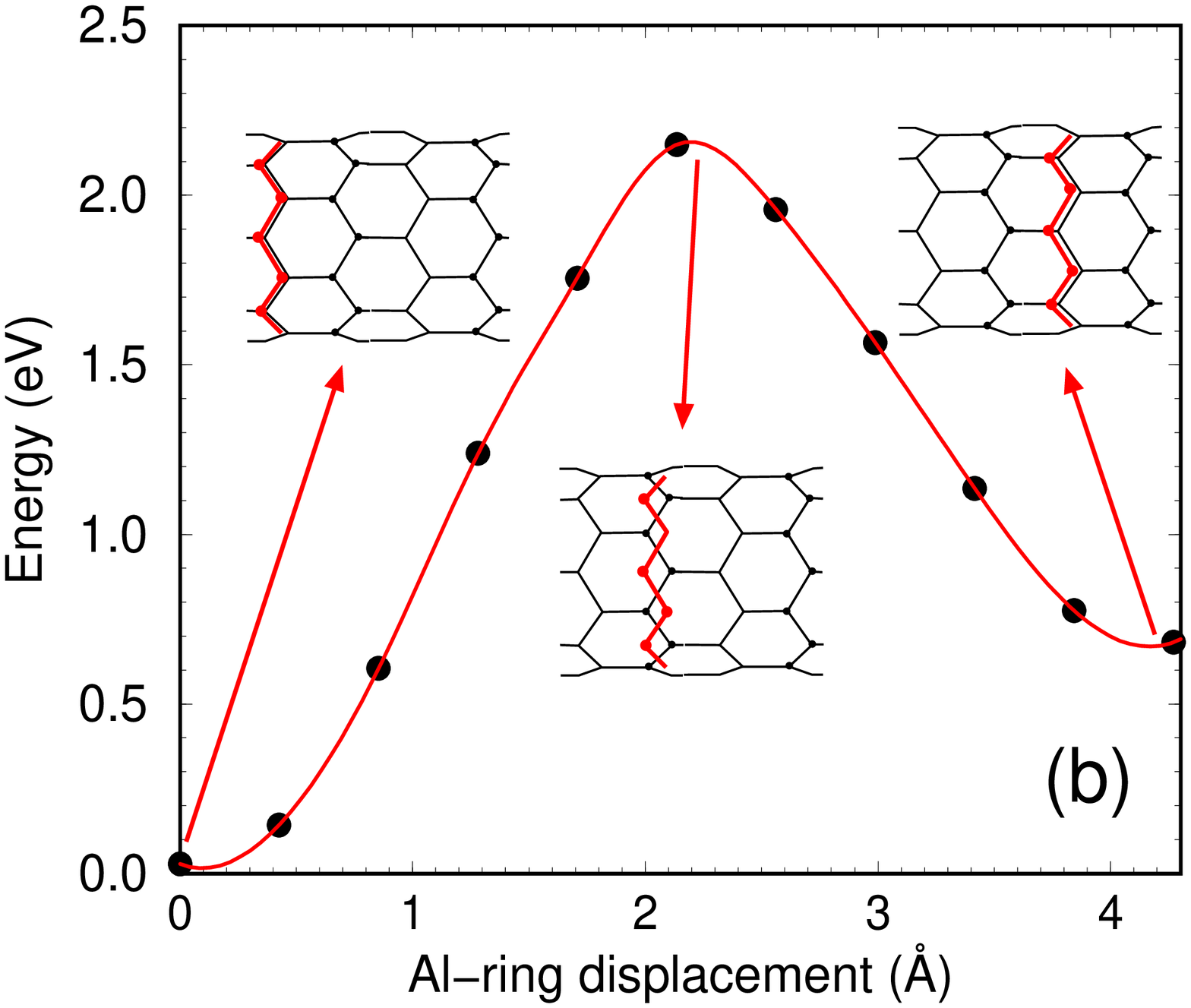}
\includegraphics[scale=0.235]{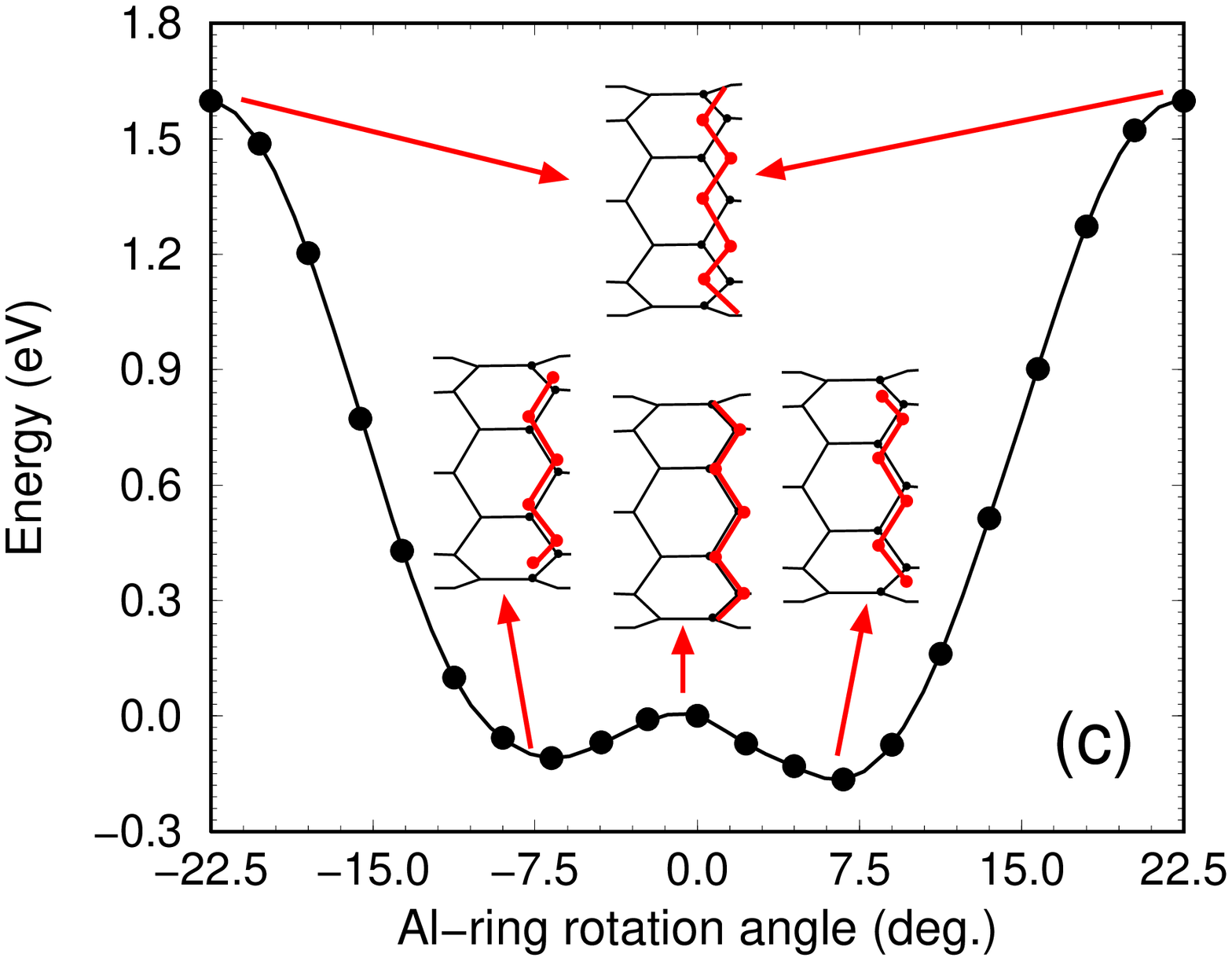}

\caption{(a) A view of the optimized structure of the Al zigzag
nanoring formed on a (8,0) SWNT.
(b) Variation of the (relative) energy with the rigid displacement
$u$ of the Al ring along the tube axis. Starting point is the
optimized structure. The insets show schematic views of nanoring
(thick gray lines) and the carbon nanotube for three particular
configurations.
(c) Variation of the energy with rigid rotation $\phi$ of the
Al ring around the nanotube. The right minimum at $\phi=7^{\circ}$
corresponds to the optimized structure; $\phi=0^{\circ}$ is the
ideal configuration with Al atoms are aligned perfectly on top
of the carbon atoms.}
\label{fig:ring}
\end{figure}

We first discuss the zigzag Al-nanoring coverage which is obtained
by placing Al atoms on top of carbon atoms (T-site), forming a
zigzag ring (See Fig~2a). This structure includes 64 C and 16 Al
atoms in the double unit cell. In this initial configuration the
Al-Al distance is 2.33 \AA{} and the angle of Al-Al-Al bond is
$\sim$137$^{\circ}$. After structure optimization, the Al-Al bond
length is increased to 2.56 \AA{}, and the Al-Al-Al bond angle is
decreased to 124$^{\circ}$, yielding the radius of the nanoring to
be 5.9 \AA{}. A side view of the optimized structure of the
Al-nanoring wrapping the (8,0) SWNT is illustrated in
Fig.~\ref{fig:ring}(a). The binding energy of the Al nanoring is
calculated to be $ 0.85 $~eV. The stability of the Al nanoring
around the nanotube can be understood from the stable structures
of planar Al monoatomic chains. Recent studies on the low
dimensional structures of metals have revealed several stable
atomic structures in one dimension
(1D).\cite{oguz1,tosatti,portal,masaku,tolla,hakki,prasen} The
first-principles calculations predicted linear chain, planar
zigzag, triangular, ladder and nonplanar dumbbell and pentagonal
structures as stable structures for Al
wires.\cite{oguz1,portal,prasen}. More interestingly, it was found
that by going from bulk to a chain structure the character of
bonding in Al wires changes and acquires
directionality.\cite{prasen} Among a number of these 1D stable
structures of Al predicted by first-principles
calculations~\cite{prasen} was the planar zigzag monoatomic chain
of Al with a bond angle 139$^{\circ}$ and bond distance 2.53~\AA.
This zigzag structure is only a local minimum on the
Born-Oppenheimer surface, and hence its binding energy (1.92 eV)
is intermediate between the binding energies of bulk and linear
structures. An energy barrier of $\sim$0.1 eV prevents the
transition from the planar zigzag structure to other relatively
more stable 1D structures.\cite{prasen} The final optimized
structure of the zigzag Al nanoring around the (8,0) nanotube has
structural parameters similar to this planar zigzag structure,
except that it is rolled on a cylinder. The SWNT initially serves
as a template in the formation of the ring structure and also
increases the stability of the ring by preventing the transitions
to other relatively more stable structures. Therefore, the Al
nanoring around the SWNT is expected to be stable at room
temperature. Interestingly, the nanoring is also stable by itself,
since the position of Al atoms do not change significantly upon
discarding the underlying carbon nanotube.

We further analyzed the SWNT-Al nanoring interaction by studying
the effect of the rigid displacement ($u$) and rotation ($\phi)$
of the nanoring around the tube axis. The variation of the energy
as a function of displacement,
$E(u)=E_T^u[Al+SWNT,u]-E_T[Al+SWNT]$ is shown in
Fig.~\ref{fig:ring}(b). Here $E_T^u[Al+SWNT,u]$ is the total
energy of the unrelaxed tube-ring system with the ring displaced
by $u$. The highest energy configuration (energetically least
favorable) corresponds to the situation where Al atoms are close
to the H-sites. For $u=c$, $E$ is 0.7~eV higher than the initial
energy with optimized structure, $E(u=0)$, since no structure
optimization was done at $u=c$. The rigid rotation of the ring in
the interval  $-22.5^{\circ} \le \phi \le 22.5^{\circ}$ is also
shown in Fig.~\ref{fig:ring}(c). The highest energy configuration
again corresponds to the case where Al atoms are close to the
H-sites (see Fig.~\ref{fig:ring}(b)). When the ring is aligned
perfectly on the top of carbon atoms at $\phi=0^{\circ}$, we
obtain a saddle point. Rotations by $\phi \sim \pm 7^{\circ}$ from
the ideal configuration ($\phi = 0^{\circ}$) results in two stable
configuration with a double minima potential separated by a
barrier of 0.2~eV. The right minimum at $\phi=7^{\circ}$
corresponds to the starting optimized structure.

\begin{figure}
\includegraphics[scale=0.45]{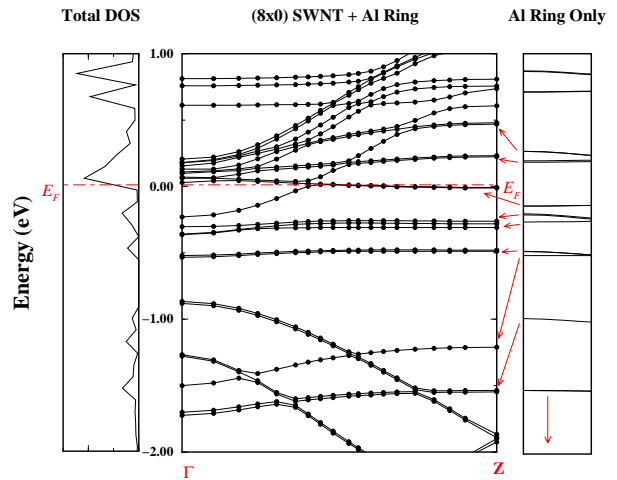}
\caption{ The energy bands along the nanotube axis of the
Al nanoring formed on a $(8,0)$ SWNT (middle panel).
The total density of states is shown in the left panel.
The right panel shows the energy levels of the bare
Al-nanoring. The zero of energy is taken at the Fermi level. }
\label{fig:band}
\end{figure}

The electronic properties of the zigzag Al nanoring system described
above (Fig.~\ref{fig:ring}(a)) is also quite interesting and may
lead to new important applications. The electronic energy bands of
the Al metal ring (without SWNT) are derived from the dispersive bands
of the flat zigzag Al chain.\cite{prasen} When the flat zigzag Al
chain is rolled into a ring, its bands are zone folded at the
$\Gamma$-point and they appear as a number of discrete energy levels
as shown in Fig.~\ref{fig:band}. For the case of the Al nanoring
wrapping the nanotube shown in Fig.~\ref{fig:ring}(a) these states
are mixed with the states of the nanotube and give rise to the bands
and density of states shown in Fig.~\ref{fig:band}. As a result of
Al-nanoring and (8,0) nanotube interaction, the combined system is a
metal. The small dispersion of the bands associated with the nanoring
are due to the small interaction between the nanorings in the supercell.
The dispersion of these bands is reduced with the increasing ring-ring
distance and eventually becomes localized states of a single ring which
dopes the empty conduction bands of the $(8,0)$ SWNT. According to
Mulliken analysis 0.15 electrons are transferred from each Al atom to
the SWNT. Most importantly, the Al ring is a conductor that incorporates
two channels with an ideal ballistic quantum conductance of $4e^2/h$.

Small radius of the metallic nanoring wrapping the carbon nanotube
may lead to interesting electromagnetic properties due to its
small radius. The magnetic field $B$ at the center of the ring can
be expressed in terms of the quantized angular momentum $L_z$ of
the electrons in the direction paralel to the tube axis:
\begin{equation}
B=\frac{\mu_0 e L_z}{4\pi m r^3},
\end{equation}
where $r$ is the radius of the nanoring. Taking the lowest
possible value for $L_z$ and $r = 5.9 $ \AA{} we estimate $B$ to
be at the order of 100 Gauss. The current in the metal ring that
can induce such a high magnetic field is comparable to the current
attained in the suspended, monoatomic gold chains.\cite{yanson}
Relatively higher magnetic fields at the order of Tesla can be
induced by higher current passing through a thick Ti based metal
coating around the SWNT, or by increasing the number of turns and
hence by forming a nanocoil. Miyamoto {\it et al.}\cite{miyamo}
have examined the chiral conductivity in bare BC$_{2}$N nanotubes.
They estimated that magnetic field of a few tenths of Tesla can be
induced at the center of the tube by assuming relaxation time of
carriers $\sim$50 times larger than that in Cu and homogeneous
chiral current density confined to the tubule wall.

Persistent currents in the nanoring can also start by sudden
application of an external magnetic field. In this way it is
possible to use a nanotube, with a ring at its end, as a local
magnetic probe at nanoscale. A superconducting ring may also be
used for Schr{\"o}dinger's cat experiments where one deals with
superposition of macroscopic quantum states.\cite{cat} The two
supercurrent quantum states (clockwise and counterclockwise flow)
sit in two separate quantum wells. It has been observed that a
weak microwave, which does not break Cooper pairs, can cause
quantum tunneling between these two macroscopic states. In this
kind of experiments the main problem is to isolate the
superconducting quantum interference device (SQUID) from the
outside (nonquantum) environment and that is why isolated carbon
nanotubes can be very useful.

\begin{figure}
\includegraphics[scale=0.4]{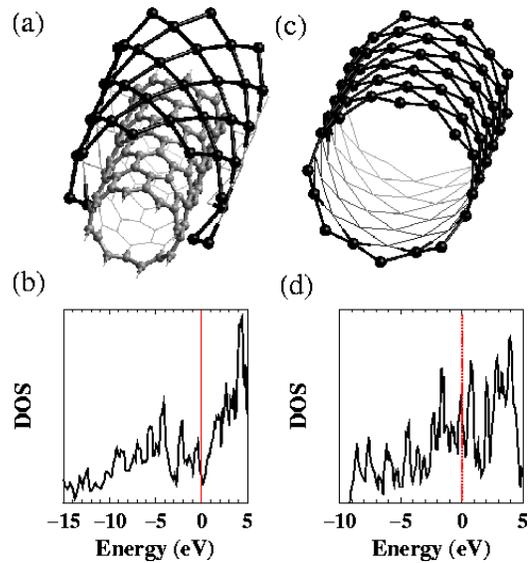}
\caption{ (a) Aluminum nanowire (dark) around SWNT (gray),
(b) Total DOS for Al nanowire + SWNT structure,
(c) Structure of Al nanowire alone, which is also stable
with smaller radius,
(d) Total DOS for the Al nanowire shown in (c).
The zero of energy is taken at $E_F$.}
\label{fig:altube}
\end{figure}

Finally we discuss another stable uniform coverage which is
obtained from sixteen Al atoms placed at the alternating T-sites
in one unit cell. Upon relaxing this structure, Al atoms move
towards the bridge sites so that the nearest neighbor distance
is 2.7 \AA{} (which is close to that of the bulk Al nearest
neighbor distance, $d_{o}= 2.8$ \AA{}) and the Al-tube distance
is 2.4 \AA. The final, stable structure can be viewed as a tubular Al
(i.e. Al nanotube), the structure of which is matched to the SWNT
with a significant tube-Al interaction (Fig.~\ref{fig:altube}).
This structure is a metal with finite density of states at $E_F$
(Fig.~\ref{fig:altube}(c)). To check whether the Al nanotube
is stable and can maintain its structure without the underlying SWNT,
we optimized the same structure without SWNT. Apart from decreasing
the radius of metal nanotube, the geometric structure preserved.
This indicates that the tubular Al is stable by itself, and the
interaction between SWNT and Al nanowire is not against the Al-Al
stability. The density of states of the bare Al nanowire without
SWNT indicates that the system is also a metal
( Fig.~\ref{fig:altube}(d)).

\section{Conclusions}

In conclusion, we show that normally Al atoms do not form a
uniform metal coverage, but they rather tend to nucleate isolated
clusters on the surface of SWNT, in agreement with the
experiments~\cite{zhang1,zhang2}. This is due to the fact that the
Al-Al interaction is stronger than that of the Al-SWNT and for most
of the decorations of Al atoms on SWNT these two interactions
compete. However, we found two special Al coverage, namely the
SWNT wrapped by the zigzag Al ring and SWNT covered by the uniform
and concentric Al nanotube. In these systems the Al-Al and
Al-nanotube interactions are not frustrated, which is the main
reason for the stability. In both cases we find significant charge
transfer from Al to SWNT, leading to metalization of the
semiconducting tubes. Al nanoring around a SWNT is of particular
interest because the states of the ring near the Fermi level can
carry ballistic current around the ring. We show that this can
give rise to persistent current and/or high magnetic fields along
the axis of the tube. Clearly this is a very promising effect for
many nano-devices applications, and provides new tools for
experiments on fundamental aspects of quantum mechanics. In spite
of the fact that Al is highly air-sensitive and easily oxidized,
it is rather taken as a prototype element in the present study. We
hope that this study will attract interest for further study to
find other elements, which form stable but air-insensitive ring
structures. We also point out technical difficulties in forming
ring structures around a SWNT. However, the recent advances in
manipulating and relocating single atoms encourage us to explore
the novel features of these nanostructures.

\begin{acknowledgments}
This work was partially supported by the NSF
under Grant No: INT01-15021 and T\"{U}B\'{I}TAK under Grant
No: TBAG-U/13(101T010).
\end{acknowledgments}


\begin{thebibliography}{99}
\bibitem{contro}
In Ref[\onlinecite{oguz2}] the binding energy of a single Al on the
$(8,0)$ tube were calculated to be $\sim$ 1 eV. This is value 0.7 eV
smaller than the present results. This is because in
Ref[\onlinecite{oguz2}] $E_{b}$ was calculated using single unit cell
incorporating significant Al-Al interaction in $E_{T}[Al]$ in
Eq.~(\ref{eq:bind}).
\end{thebibliography}
\end{document}